\documentclass[prb,
superscriptaddress,showpacs,amsmath,amssymb]{revtex4}

\usepackage{graphicx}

\begin{document}

\title{Black hole and Hawking radiation by type-II Weyl fermions}

\author{G.E.~Volovik}
\affiliation{Low Temperature Laboratory, Aalto University,  P.O. Box 15100, FI-00076 Aalto, Finland}
\affiliation{Landau Institute for Theoretical Physics, acad. Semyonov av., 1a, 142432,
Chernogolovka, Russia}

\date{\today}

\begin{abstract}
The type-II Weyl and type-II Dirac fermions may emerge  behind the event horizon of black holes.
Correspondingly the black hole can be simulated by creation of the region with overtilted Weyl or Dirac cones. The filling of the electronic states inside the "black hole" is accompanied by Hawking radiation. 
The Hawking temperature in the Weyl semimetals can reach the room temperature, if the black hole region is sufficiently small, and thus the effective gravity at the horizon is large.

\end{abstract} 


\maketitle

\section{Introduction}
\label{sec:Intro}

 The elementary particles of Standard Model (quarks and leptons)  are originally the fermions obeying the Weyl equation.\cite{Weyl1929}  In condensed matter the Weyl fermions live in the vicinity of the Weyl point. The Weyl point is the exceptional point of level crossing in the space of 3 parameters,\cite{NeumannWigner1929} in this case  these are three components of the momentum $p_x$, $p_y$ and $p_z$  of the 3D energy spectrum. This crossing point is topologically protected.\cite{Novikov1981} It represents the monopole in the Berry phase flux,\cite{Volovik1987,Volovik2003} which  is described by the momentum-space topological invariant,  the Chern number $N_3=\pm 1$.   The Weyl point with higher Chern numbers, $|N_3| > 1$,  
 describes the higher order touching of two energy bands.\cite{VolovikKonyshev1988}   
Weyl fermions exist in chiral superfluid $^3$He-A, where the corresponding effects --  the chiral anomaly\cite{Bevan1997,Volovik2003} and chiral magnetic effect\cite{Krusius1998,Volovik1998} -- have been experimentally observed. They Weyl fermions are also typical in topological 
semimetals.\cite{Herring1937,Abrikosov1971,Abrikosov1972,NielsenNinomiya1983,Burkov2011a,Burkov2011b,Weng2015,Huang2015,Lv2015,Xu2015,Lu2015}

Recently  the so-called type-II Weyl points attracted attention in condensed matter.\cite{Soluyanov2015,YongXu2015,Chang2016,Autes2016,Xu2016,Jiang2016,Beenakker2016,
HuangMcCormock2016,XuWang2016,DengWan2016,LiangHuang2016,AliXiong2014,
WuJo2016,XuZahng2015,YuYao2016,Udagawa2016,ZyuzinTiwari2016}
In relativistic theories the type-II Weyl fermions also appear. They emerge in particular in the vacuum of the real (Majorana) fermions.\cite{VolovikZubkov2014} In general the $2 \times 2$ Hamiltonian describing the Weyl fermions in the vicinity of the topologically protected Weyl point at ${\bf p}^{(0)}$ has the following general form
\begin{equation}
H= e_k^j(p_j-p^{(0)}_j) \hat\sigma^k  + e_0^j(p_j-p^{(0)}_j)+ ... \,\,.
\label{HamiltonianGeneral} 
\end{equation}
This expansion demonstrates that position ${\bf p}^{(0)}$ of the Weyl point plays the role of the effective dynamical $U(1)$ gauge field, while $e_k^j$
and $e_0^j$ are the emergent tetrad fields. The energy spectrum of this Hamiltonian is determined by the ratio between the two terms in the rhs of Eq.(\ref{HamiltonianGeneral} ), i.e. on the parameter 
$|e^j_0 [e^{-1}]_j^k| $.\cite{VolovikZubkov2014} When
$|e^j_0 [e^{-1}]_j^k|  < 1$ one has the conventional Weyl point with the tilted Weyl cone. At
 $|e^j_0 [e^{-1}]_j^k| > 1$ the cone is overtilted and two Fermi surfaces appear, which touch each other at the Weyl point. This is the regime of the type-II Weyl point. 

 \section{Type-II Weyl fermions behind the event horizon}
 \label{sec:BH}

The type-II Weyl point appears 
also behind the event horizon.\cite{HuhtalaVolovik2002,Volovik2003,Volovik2016}
 In
general relativity the stationary metric, which is valid both outside and inside the black hole horizon, is provided in particular by the Painlev\'e-Gullstrand spacetime.\cite{Painleve} The line
element of the Painlev\'e-Gullstrand metric is equivalent to the so-called 
acoustic metric:\cite{unruh1,unruh2,Kraus1994} 
\begin{equation}
 ds^2= g_{\mu\nu}dx^\mu dx^\nu=- c^2dt^2+ (d{\bf r}-{\bf v}dt)^2 = - (c^2-
v^2)dt^2- 2{\bf v}d{\bf r}dt+d{\bf r}^2 \,.
\label{Painleve}
\end{equation} 
This is stationary but not static metric, which is expressed in terms of the velocity field ${\bf v}({\bf r})$ describing the frame dragging in the gravitational field. 
For the spherical black hole the velocity field is radial:
\begin{equation}
{\bf v}({\bf r})= - \hat{\bf r}c\sqrt{\frac{r_h}{r}} ~,~r_h=\frac{2MG}{c^2} \,.
\label{VelocityField}
\end{equation} 
Here $M$ is the mass of the black hole; $r_h$ is the radius of the horizon;
 $G$ is the Newton
gravitational constant. The minus sign in
Eq.(\ref{VelocityField}) gives the metric for the black hole, while
the plus sign would characterize the white hole. 
 The corresponding tetrad fields are $e_k^j=c\delta_k^j$ and $e_0^j=v^j$.\cite{Doran2000} 
 Behind the black hole horizon  the drag velocity exceeds the speed of light, ${\bf v}^2({\bf r}) > c^2$.

\begin{figure}
 \includegraphics[width=0.5\textwidth]{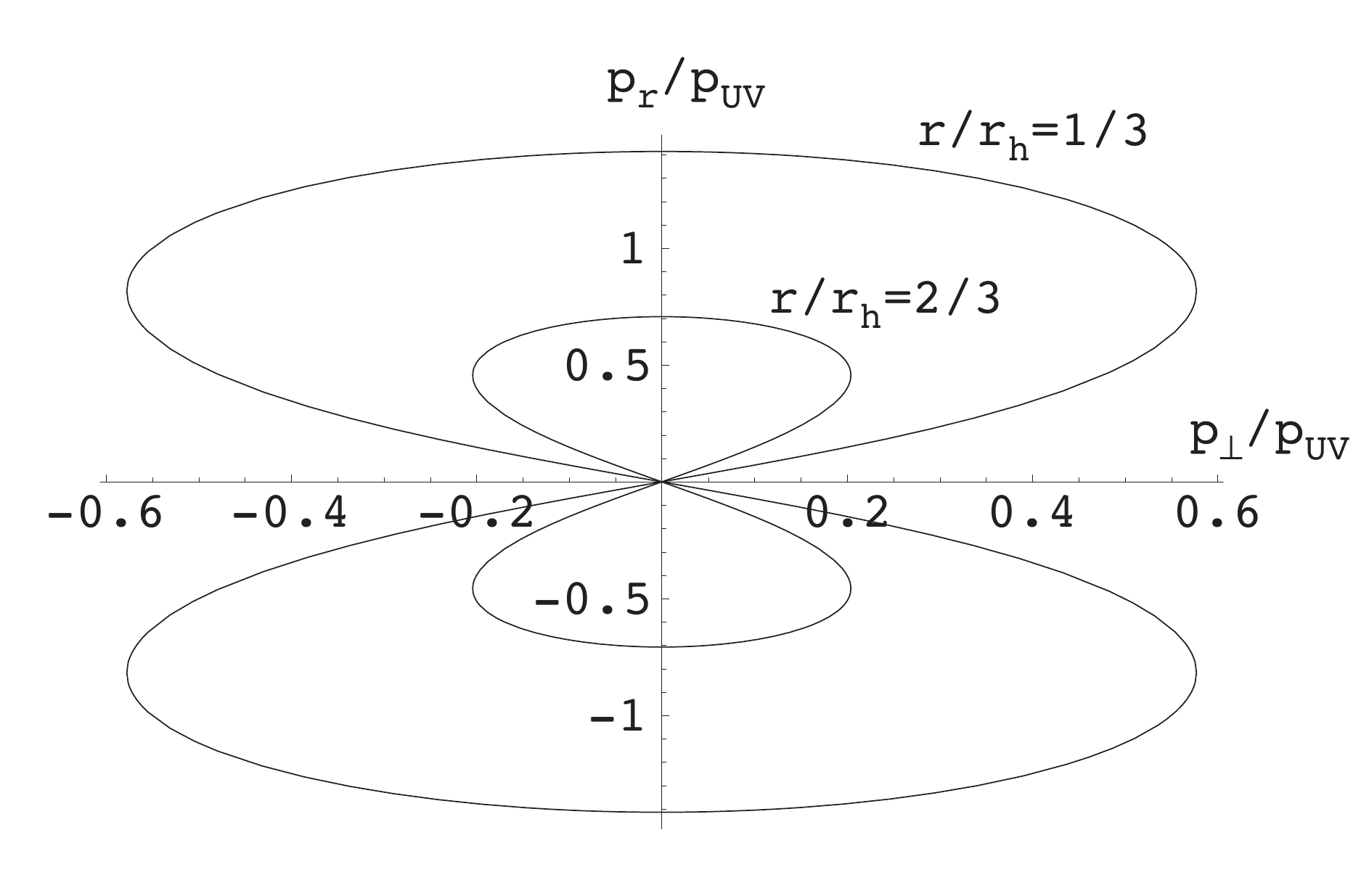}
 \caption{Fermi surfaces of the relativistic  type-II Weyl fermions behind the black hole horizon  (from Ref.\cite{HuhtalaVolovik2002}). The Fermi surfaces depend on the position inside the horizon and are shown for $r=r_h/3$ and  $r=2r_h/3$ where $r$ is the distance from the center of the spherical black hole. 
Here $p_r$ is the radial component of the momentum ${\bf p}$; $p_\perp  =\sqrt{ p^2 - p_r^2}$; and $p_{\rm UV}= E_{\rm UV}/c$ is the ultraviolet cut-off at which the Lorentz invariance is violated.
}
 \label{FermiSurfaceBH}
\end{figure}

The energy spectrum of the relativistic fermions living in the Painlev\'e-Gullstrand spacetime is given by the contravariant metric tensor, $g^{\mu\nu}p_\mu p_\nu=0$.   Near the horizon of the black hole and inside the horizon the particle masses can be neglected, and one can consider the left handed and right handed fermions independently. The corresponding Hamiltonian for the fundamental Weyl fermions interacting with the tetrad field corresponding to the Painlev\'e-Gullstrand metric has the form:\cite{HuhtalaVolovik2002}
 \begin{equation}
 H=  
\pm c{\mbox{\boldmath$\sigma$}} \cdot{\bf p}  - p_r v(r) + \frac{c^2 p^2}{E_{\rm UV}}\,\,, \,\, v(r)=c\sqrt{\frac{r_h}{r}} \,.
 \label{HamiltonianBH}
 \end{equation}
Here the plus and minus signs correspond to the right handed and left handed fermions respectively; $p_r$ is the radial momentum of fermions. The second term in the rhs of (\ref{HamiltonianBH}) is the Doppler shift due to the drag velocity, ${\bf p}\cdot {\bf v}({\bf r})$. The third term is the added nonlinear dispersion. The latter is natural for the condensed matter systems. In particle physics the  parameter $E_{\rm UV}$ is the ultraviolet (UV) energy scale at which the violation of the Lorentz invariance can be expected. 
(Note that the current experimental bounds on Lorentz-violating modifcations suggest that in our vacuum the UV scale is much higher than the Planck energy scale.\cite{Kostelecky2011}) 
However, the nonlinear term can arise in effective Hamiltonian $H=G^{-1}(\omega=0,{\bf p})$ even without violation of Lorentz invariance on fundamental level:\cite{Volovik2010} the Green's function $G(\omega,{\bf p})$ may still be relativistic invariant, while the  Lorentz invariance of the Hamiltonian is violated due to the existence of the heat bath reference frame outside the black hole.

In the heat bath reference frame the spectrum of Weyl fermions behind the horizon, where ${\bf v}^2({\bf r}) > c^2$, has the following properties. The negative energy states of the originally positive branch
 \begin{equation}
 E_+=  
 (c - v(r))p_r  + \frac{c^2 p^2}{E_{\rm UV}}  \,,
 \label{PositiveBranch}
 \end{equation}
form the Fermi surface. Together with the Fermi surface of holes (anti-particles) one has the pair of the Fermi surfaces  attached to the Weyl point. This  corresponds to the type-II Weyl point. The Fermi surfaces are finite if the $p^2$ term is taken into account. These Fermi surfaces are shown 
in Fig. \ref{FermiSurfaceBH} at two positions inside the black hole: at $r=r_h/3$ and $r=2r_h/3$. 
Close to the horizon, where $r_h - r \ll r_h$, the Fermi surfaces are concentrated in the region
 $|p_r|<p_{\rm UV}(v(r)-c)/c \ll p_{\rm UV}$, and
the nonlinear $p^2$ term can be considered as a small perturbation.

 \section{Artificial black hole with type-II Weyl fermions}
\label{sec:ArtBH}
 
\begin{figure}
 \includegraphics[width=0.8\textwidth]{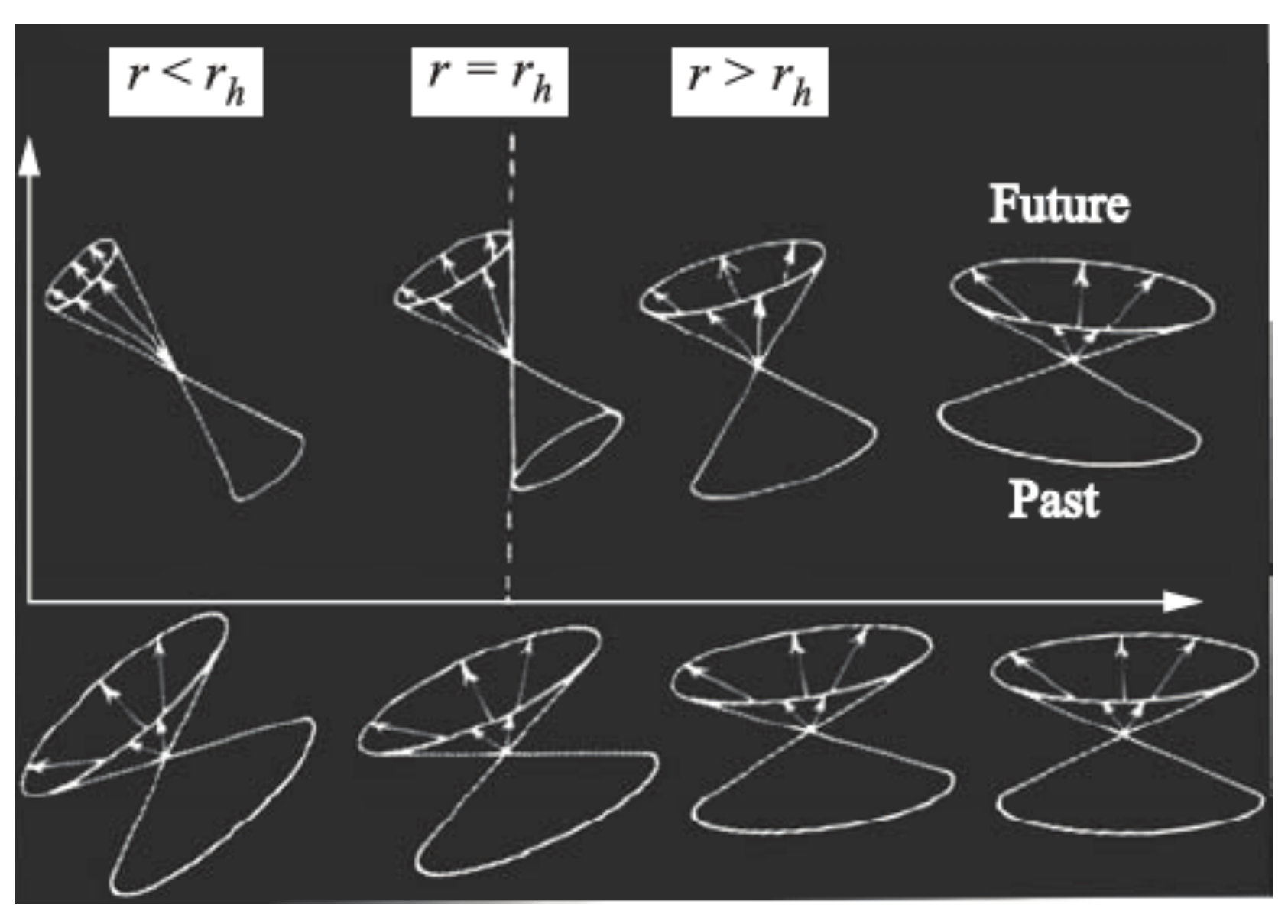}
 \caption{Illustration of the black hole horizon at $r=r_h$ and of the artificial event horizon at the interface between the type-I ($r>r_h$) and type-II ($0\leq r<r_h$) Weyl semimetals. 
 ({\it top}): The  light cone  for relativistic Weyl and Dirac fermions, $g_{\mu\nu}x^\mu x^\nu=0$,  outside and inside the black hole in the Painlev\'e-Gullstrand space-time.  At $r<r_h$ the light cone is overtilted and relativistic fermions with linear spectrum are locked inside the horizon. 
 ({\it bottom}): The corresponding Dirac cone in the energy spectrum of relativistic Weyl fermions is described by the contravariant metric tensor, $g^{\mu\nu}p_\mu p_\nu=0$.  
At $r=r_h$ the cone touches the zero energy level. At $r<r_h$ the cone is overtilted and two Fermi surfaces are formed, which are finite if the nonlinear corrections are added, see Fig. \ref{FermiSurfaceBH}.  In Weyl semimetals, the Dirac cone in the energy-momentum space, $g^{\mu\nu}p_\mu p_\nu=0$, is the primary quantity, where ${\bf p}$ is counted from the position of the Weyl point. 
The corresponding covariant metric tensor, $g_{\mu\nu}x^\mu x^\nu=0$, describes the effective space-time for Weyl fermions.  This metric experiences the event horizon at the boundary between  the type-I ($r>r_h$) and type-II ($0\leq r<r_h$) Weyl semimetals.
In the equilibrium state, when all the negative energy levels inside the horizon are occupied, the black hole is not radiating. However, just after formation of such black hole the process of filling  of the negative energy states includes the channel corresponding to the thermal Hawking radiation with temperature in Eq.(\ref{HawkingT}).
 }
 \label{BH}
\end{figure}

The behavior of the energy spectrum behind the event horizon has one to one correspondence with the energy spectrum of the type-II Weyl fermions. That is why one can invert the consideration in Sec.\ref{sec:BH} and construct the analog of the black hole horizon starting with the fermionic spectrum of Weyl semimetals. The Hamiltonian (\ref{HamiltonianGeneral}) determines the effective tetrad field, which in turn gives rise to the effective contravariant metric,  
 $g^{\mu\nu}=e^\mu_a e^\nu_b \eta^{ab}$. The latter describes the conical spectrum of Weyl quasiparticles, $g^{\mu\nu}(p_\mu-p^{(0)}_\mu)(p_\nu-p^{(0)}_\nu)=0$. The corresponding covariant metric $g_{\mu\nu}$ determines the effective space-time, in which the quasiparticles live, and describes the behavior  of the effective "light cone". 

 Let us now consider the inhomogeneous semimetal, where the transition
between the type-I and type-II Weyl points takes place at some surface. Let us discuss, for example, a spherical surface of radius $r_h$, with 
type-II Weyl fermions inside the sphere, see Fig. \ref{BH} ({\it bottom}). Then the Hamiltonian has the form of Eq.(\ref{HamiltonianBH}) with $v(r_h)=c$, and the effective space-time is determined by covariant metric $g_{\mu\nu}$  in Eqs.(\ref{Painleve})-(\ref{VelocityField}). At $r<r_h$ the effective light cone is overtilted, which simulates the interior of the black hole in the 
Painlev\'e-Gullstrand space-time in Fig. \ref{BH} ({\it top}).

The formed black hole horizon has the following property. It is fully stationary in equilibrium, when all the negative energy states in the Fermi surfaces  inside the horizon are occupied. Such equilibrium  black hole is not radiating, as is probably distinct from the black hole in the  fundamental  Einstein theory of gravity. However, at the first moment of creation of the black hole analog, the system is not in the equilibrium state:  the negative energy states of the former positive branch $E_+$ are empty, while the positive energy states of the former negative branch $E_-$  are occupied. The initial stage of equilibration --  the filling of the negative energy states by the fermions occupying the positive energy states --  corresponds to creation of the particle-hole pairs at the horizon and thus simulates the Hawking radiation. The corresponding  Hawking  temperature is determined by effective gravitational field at the horizon: 
  \begin{equation}
 T_{\rm H}= \frac{\hbar}{2\pi} \left(\frac{dv}{dr}\right)_{r=r_h} \,.
 \label{HawkingT}
 \end{equation} 
For the sufficiently small 
black hole regions the gradient of the system parameter $v$ can be large, and $T_{\rm H}$ may reach the room temperature.

\section{Conclusion}
\label{sec:Conclusion}
 
Here we considered the mechanism of formation of the artificial event horizon, which is distinct from the traditional dynamic mechanism based on the supercritical
 flow behind the horizon.\cite{unruh1,unruh2,Barcelo2001,Volovik2003,Lahav2010} 
The instrumental is the behavior of the fermions in the vicinity of the type-II Weyl points in the topological semimetals, which is similar to the behavior of the relativistic Weyl fermions behind the horizon in  the Painlev\'e-Gullstrand space-time. The analog of the event horizon is formed at the interface between the type-I and type-II Weyl semimetals. In equilibrium the black hole analog is not radiating, but the equilibrization after the formation of the black hole includes the process, which is analogous to the Hawking radiation.
As distinct from the other suggested  realizations of the Hawking radiation, the Hawking temperature in the Weyl semimetals can reach the room temperature.

\section*{\hspace*{-4.5mm}ACKNOWLEDGMENTS}
I acknowledge financial support from the ERC,  Advanced Grant project TOPVAC.

\end{document}